\documentclass[10pt,twocolumn,twoside]{IEEEtran}
\makeatletter
\long\def\@makecaption#1#2{\ifx\@captype\@IEEEtablestring%
	\footnotesize\begin{center}{\normalfont\footnotesize #1}\\
		{\normalfont\footnotesize\scshape #2}\end{center}%
	\@IEEEtablecaptionsepspace
	\else
	\@IEEEfigurecaptionsepspace
	\setbox\@tempboxa\hbox{\normalfont\footnotesize {#1.}~~ #2}%
	\ifdim \wd\@tempboxa >\hsize%
	\setbox\@tempboxa\hbox{\normalfont\footnotesize {#1.}~~ }%
	\parbox[t]{\hsize}{\normalfont\footnotesize \noindent\unhbox\@tempboxa#2}%
	\else
	\hbox to\hsize{\normalfont\footnotesize\hfil\box\@tempboxa\hfil}\fi\fi}
\makeatother

\usepackage{cite}
\usepackage[dvips]{graphicx}
\usepackage[cmex10]{amsmath}
\usepackage{amssymb}
\usepackage[ruled,linesnumbered]{algorithm2e}
\usepackage{array}
\usepackage{mdwmath}
\usepackage{mdwtab}
\usepackage{eqparbox}
\usepackage{multirow}
\usepackage[tight,footnotesize]{subfigure}
\usepackage{stfloats}
\usepackage[hidelinks]{hyperref}
\usepackage{rotating,setspace}
\begin{document}

\title{Towards End-to-End Synthetic Speech Detection}
\author
{
	Guang~Hua, \emph{Member, IEEE},
	Andrew Beng Jin Teoh, \IEEEmembership{Senior Member, IEEE}, and
	Haijian Zhang, \IEEEmembership{Member, IEEE}
	\thanks{This work was supported by the 2020\---2021 International Scholar Exchange Fellowship (ISEF) Program at the Chey Institute for Advanced Studies, South Korea. \emph{(Corresponding Author: Andrew Beng Jin Teoh)}}
	\thanks{G. Hua and H. Zhang are with the School of Electronic Information, Wuhan University, Wuhan 430072, China (e-mail: ghua@whu.edu.cn; haijian.zhang@whu.edu.cn).}
	\thanks{A. B. J. Teoh is with the School of Electrical and Electronic Engineering, College of Engineering, Yonsei University, Seoul 120749, South Korea (e-mail: bjteoh@yonsei.ac.kr).}
	
}

\markboth{IEEE Signal Processing Letters}
{HUA \MakeLowercase{\emph{et al.}}, \AmS-\LaTeX}
\maketitle

\begin{abstract}
The constant Q transform (CQT) has been shown to be one of the most effective speech signal pre-transforms to facilitate synthetic speech detection, followed by either hand-crafted (subband) constant Q cepstral coefficient (CQCC) feature extraction and a back-end binary classifier, or a deep neural network (DNN) directly for further feature extraction and classification. Despite the rich literature on such a pipeline, we show in this paper that the pre-transform and hand-crafted features could simply be replaced by end-to-end DNNs. Specifically, we experimentally verify that by only using standard components, a light-weight neural network could outperform the state-of-the-art methods for the ASVspoof2019 challenge. The proposed model is termed Time-domain Synthetic Speech Detection Net (TSSDNet), having ResNet- or Inception-style structures. We further demonstrate that the proposed models also have attractive generalization capability. Trained on ASVspoof2019, they could achieve promising detection performance when tested on disjoint ASVspoof2015, significantly better than the existing cross-dataset results. This paper reveals the great potential of end-to-end DNNs for synthetic speech detection, without hand-crafted features.
\end{abstract}

\begin{IEEEkeywords}
Synthetic speech detection, speech forensics, ASVspoof2019, ASVspoof2015, cross-dataset testing, end-to-end.
\end{IEEEkeywords}

\IEEEpeerreviewmaketitle

\section{Introduction}
\IEEEPARstart{T}{he} success of deep learning technology has shifted the paradigm of speech synthesis from the classic hidden Markov model based  framework \cite{2013_Tokuda_ProcIEEE_HMM} to neural speech synthesis. Equipped with powerful deep neural network (DNN) architectures  e.g., \cite{2021_Ren_ICLR_FastSpeech2}, and fueled by massive training data, today's text-to-speech (TTS) systems could synthesize high quality speech that is hard to be distinguished from human voices. Despite the multitude of benefits, these advances have also improved the quality of voice spoofing attacks, including voice conversion \cite{2017_Tian_TASLP_VC}, impersonation \cite{2018_Gao_ICASSP_Impersonation}, cloning \cite{2018_Arik_NIPS_Voice_Clone}, etc., posing new challenges to synthetic speech detection.

For nearly a decade, the combination of a front-end feature extractor and a back-end binary classifier is the \emph{de facto} framework for synthetic speech detection. Within this framework,  an overwhelming majority of the existing works have focused on the development of hand-crafted front-end features, including fundamental frequency, power spectrum, octave spectrum, linear
frequency cepstral coefficient (LFCC), mel-frequency cepstral coefficient (MFCC), cepstral mean and variance (CMVN), cochlear filter cepstral coefficient (CFCC), filter bank based cepstral coefficient, linear prediction cepstral coefficient (LPCC), modified group delay (MGD), relative phase shift (RPS),  constant Q cepstral coefficient (CQCC), and many of their variations and combinations \cite{2015_Sanchez_TIFS_Det,2015_Sahidullah_INTERSPEECH_Det,2016_Sara_SC_Phase_Det,2017_Patel_JSTSP_Det,2017_Patel_JSTSP_IF_Det,2017_Paul_JSTSP_Det,2017_Wang_JSTSP_Phase_Det,2017_Todisco_CSL_CQT_Det,2018_Pal_CSL_hand_Det,2019_Yang_TASLP_Octave_Det}. Usually, one or a few of these features are used to train a Gaussian mixture model  (GMM) or a support vector machine (SVM) for classification. Taking the advantage of DNNs in classification tasks, multilayer perceptron (MLP) and convolutional neural network (CNN) based classifiers have been used to replace the conventional back-end classifiers \cite{2016_Tian_APSIPA_CNN_Det, 2017_Muck_TASLP_Hand_Det,2018_Yu_TNNLS_Acoustic_Det,2019_Yang_TASLP_Octave_Det,2020_Yang_TIFS_Subband_Det,2017_Zhang_JSTSP_DL_Det, 2019_Fuse_All,2017_Chen_INTERSPEECH_ResNet_Det_Replay}. On the other side, DNN structures have also been used at the front-end to facilitate feature extraction \cite{2016_Qian_SC_Deep_Det,2017_Chen_INTERSPEECH_ResNet_Det_Replay,2020_Adiban_CSL_Siamese_Det,2017_Qian_TASLP_Deep_Det}, followed by conventional classifiers. DNNs can also work across the front- and back-end, with pre-transformed features as input \cite{2019_Ensenble,2019_SE_Res, 2019_VGG,2019_STC,2021_Li_ICASSP_Res2Net_Det}.


\begin{figure}[!t]
	\centering
	\includegraphics[width=2.8in]{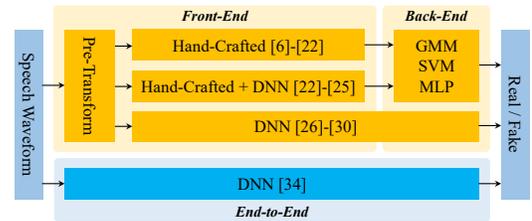}
	\caption{Relationship between the existing front-end$\rightarrow$back-end pipeline and the proposed end-to-end framework for synthetic speech detection.}
	\label{taxonomy}
\end{figure}

Among hand-crafted features, CQCC has been found to be the best choice, which is also the baseline feature in the ASVspoof2019 challenge \cite{ASVspoof2019_short}. Recently, Yang \emph{et al.} developed a set of subband CQCC features for better detection performance \cite{2020_Yang_TIFS_Subband_Det}. Subsequently, Das \emph{et al.}\cite{2019_Fuse_All} further fused $8$ hand-crafted features, followed by an MLP classifier. For deep learning based approach, Lavrentyeva \emph{et al.} \cite{2019_STC} proposed the use of FFT, LFCC, and CMVN, followed by a CNN for classification, while using CQT as model input, Li \emph{et al.} \cite{2021_Li_ICASSP_Res2Net_Det} incorporated the so called Res2Net structure and squeeze-and-excitation (SE) block. With score level fusion, Lavrentyeva \emph{et al.} \cite{2019_STC} and  Li \emph{et al.} \cite{2021_Li_ICASSP_Res2Net_Det} have achieved the state-of-the-art performance on ASVspoof2019 dataset. 

Based on the above overview, the existing mainstream workflow for synthetic speech detection is summarized in the brown blocks of Fig. \ref{taxonomy}. It can be seen that a time-frequency transform (e.g., CQT) of the speech waveform before  hand-crafted feature extraction (e.g., CQCC), or before feeding the data into a DNN, has become an implicit standard routine in the existing works. However, since DNNs are best known for their excellent capability of feature extraction, there naturally arises a question of whether it is necessary to apply these pre-transforms. In fact, these transforms usually discard some information about the observed speech signal. For example, the CQT feature, more precisely the log power spectrum of the CQT \cite{2021_Li_ICASSP_Res2Net_Det}, does not have the phase information of the signal. To further generate the CQCC, even more information will be discarded \cite{2020_Yang_TIFS_Subband_Det}. From hand-crafted feature engineering point of view, a good feature captures discriminative information between classes and is also compact in size, but the same principle may not apply to the DNN regime.

In this paper, we show that the pre-transforms, as well as the hand-crafted features, are in fact not a must for DNN based synthetic speech detection. Despite the rich hand-crafted features, we experimentally verify that via the use of standard DNN structures, an end-to-end light-weight neural network with mere speech waveform could achieve even better results. Our proposal is motivated by recent works analyzing raw-waveform based DNNs \cite{End2End_2019_Interspeech} and the attempt of applying end-to-end DNNs to related speech processing tasks, e.g., speech separation \cite{End2End_2019_ConvTasNet}. The proposed model is thus termed as Time-domain Synthetic Speech Detection Net (TSSDNet). We note that the first work on end-to-end synthetic speech detection was probably carried out by Muckenhirn \emph{et al.} \cite{2017_Muck_IJCB_E2E_Det}, in which a basic feedforward sequential CNN was used. It was tested on older datasets, not achieving the state-of-the-art results. In our design of the TSSDNet, two types of advanced CNN structures are considered, including ResNet-style skip connection with $1 \times 1$ kernels \cite{2015_ResNet} and Inception-style parallel convolutions \cite{2015_Inception}, respectively. We demonstrate that via proper training, the proposed networks outperform the state-of-the-art hand-crafted feature based detectors as well as DNN based ones on the challenging ASVspoof2019 dataset \cite{ASVspoof2019_short}. To analyze practical merits of the proposed methods, we further perform a cross-dataset evaluation between ASVspoof2019 and ASVspoof2015\cite{ASVspoof2015}\footnote{The ASVspoof2017 dataset is not considered in this paper because it only has replay attack. Although replay attack is seen to be considered together with synthesis attack, the underlying mechanism is very different. The physical access portion of ASVspoof2019 is also excluded for the same reason.}, demonstrating their promising generalization capability.

\begin{figure}[!t]
	\centering
	\subfigure[ResNet style, Res-TSSDNet.]{\includegraphics[width=2.9in]{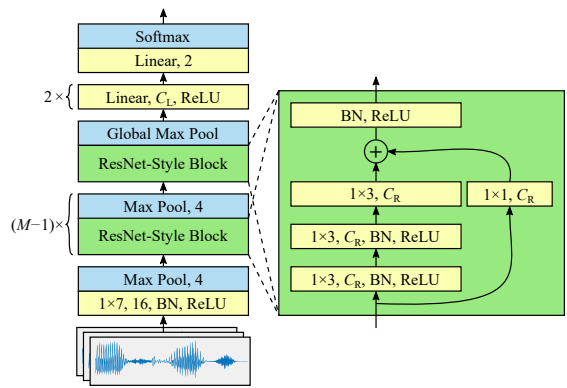}}\\
	\subfigure[Inception style, Inc-TSSDNet.]{\includegraphics[width=2.65in]{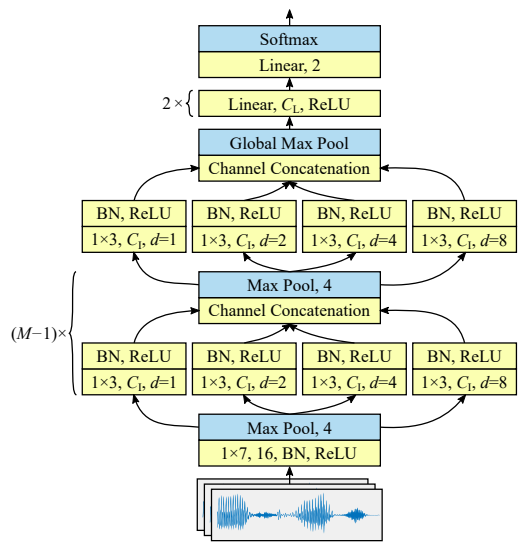}}
	\caption{Structures of the proposed models, where all the conv layers apply ``SAME'' padding, and in all local max pooling layers, stride$=$kernel size. $M$: number of stacked ResNet- and Inception-Style modules. $C_\text{R}$: number of channels in Res-TSSDNet. $C_\text{I}$: number of channels in Inc-TSSDNet.}
	\label{models}
\end{figure}

\section{The Proposed Models}
In many deep learning tasks such as object recognition or semantic understanding, it has been found that generally the deeper the network, the better the performance \cite{2015_ResNet,2015_Inception,2015_VGG}. However, in synthetic speech detection, the critical feature is the artifact left behind data forgery, which may not contain any semantic information. Since deeper features are more towards higher level semantic information, which may not be suitable to represent the subtle forgery artifacts, we hypothesize that the network for synthetic speech detection should be relatively shallower. Grabbing the essence of the popular ResNet \cite{2015_ResNet} and Inception network \cite{2015_Inception}, the proposed end-to-end TSSDNets are designed as follows.

\subsection{Model Structure}
The proposed Res-TSSDNet and Inc-TSSDNet are depicted in Fig. \ref{models} (a) and (b), respectively, which share the same first layer, $3$ final fully-connected linear layers, and global max pooling before the linear layers. The ResNet-style and Inception-style blocks are repeated for $M$ times, respectively, and batch normalization (BN) is applied in both networks. $C_\text{R}$ and $C_\text{I}$ denote the number of channels in the corresponding modules, which may vary across layers. Noticeably, to increase the receptive field and control model complexity, dilated convolutions \cite{2016_Dilated_conv} with dilation $d$ are incorporated in the Inc-TSSDNet, which is different from the original Inception network \cite{2015_Inception}. All the convolution layers apply ``SAME'' padding with $\text{stride}=1$, while for the pooling layers the stride equals to the corresponding kernel size.

\subsection{Training Strategy}
\subsubsection{Data Preparation}
Normally, the training data contain raw speech recordings with varied durations. To align the training data, we adopt the treatment in \cite{2021_Li_ICASSP_Res2Net_Det}. In \cite{2021_Li_ICASSP_Res2Net_Det}, the training examples are truncated or repeated until the duration is $6.4$ seconds to generate the CQT feature, while here we keep every example with $6$ seconds, with the default $16$ kHz sample rate. These $6$-second examples are then directly fed into the networks for end-to-end training. Since all convolution layers have ``SAME'' padding, the length of feature vector ($9.6\times 10^4$ at input) is reduced solely by the pooling layers. Note that hand-crafted feature based method, e.g., CQCC \cite{2020_Yang_TIFS_Subband_Det}, is insensitive to the length of recording since all time slices contribute to classifier training. Batch size is set to $32$. .

\subsubsection{Weighted Cross-Entropy Loss} 
Considering the fact that in general data-driven media content forgery detection tasks, the number of genuine examples is usually much less than the number of fake ones, we apply weighted cross-entropy (WCE) loss during the training phase to cope with data imbalance. Let $\{x_i, y_i\}$ compose the labeled training set, where $\forall i$, label $y_i\in\{0,1\}$, then the WCE loss is given by
\begin{equation}\label{WCE}
	{\mathop{\rm WCE}\nolimits} \left( {{\bf{z}},{y_i}} \right) =  - {w_{{y_i}}}\log \left( {{z_{{y_i}}}} \right),
\end{equation}
where $\mathbf{z}=[z_0, z_1]$ contains the softmax probabilities of the $2$ classes, and $w_{y_i}$ is the inverse ratio of label $y_i$ in the training set. For all the training processes, we use the Adam \cite{2017_Adam} optimizer and default settings. Exponential learning rate decay with a multiplicative factor of $0.95$ is adopted. The model yielding the lowest equal error rate (EER) on development set within $100$ epochs is selected for evaluation.

\subsubsection{Mixup Regularization} 
For practical forensic merits, the trained model is expected to generalize to unseen attacks, and the ASVspoof datasets have been specially designed for this purpose. In this paper, we consider the mixup regularization \cite{2018_mixup} as a booster to further improve the generalization capability. Specifically, it uses a set of mixed examples and labels, instead of the original set, to train the network, i.e.,
\begin{equation}
	\tilde{x}_i = \lambda x_i + (1-\lambda) x_j,\quad \tilde{y}_i = \lambda y_i + (1-\lambda) y_j,
\end{equation}
where $\{x_i, y_i\}$ and $\{x_j, y_j\}$ are two randomly selected training pairs, $\lambda\sim \mathop{\rm Beta}(\alpha, \alpha)$, and $\alpha\in(0, \infty)$ is a hyperparameter. The implementation of the mixup regularization could be carried out via the following equivalent loss function,
\begin{equation}
{{\mathop{\rm CE}} _{{\rm{mixup}}}}\left( {\tilde{\bf{z}},{y_i},{{y}_j}} \right) = \lambda {\mathop{\rm CE}} (\tilde{\bf{z}},{y_i}) + (1 - \lambda ){\mathop{\rm CE}} (\tilde{\bf{z}},{{y}_j}), 
\end{equation}
where $\tilde{\bf{z}}$ contains the softmax probabilities from mixed examples, and ${\mathop{\rm CE}}(\cdot, \cdot)$ is the standard cross-entropy (CE) loss, equivalent to setting $w_0=w_1$ in (\ref{WCE}).

\section{Results}
We first present the main results obtained by the proposed networks in comparison with the benchmark and the state-of-the-art solutions on the latest ASVspoof2019 dataset. We then perform ablation study, followed by cross-dataset evaluation on the ASVspoof2015 dataset. All the results are generated using a single GeForce GTX 1080 or 1080Ti GPU. PyTorch implementations of the proposed TSSDNets are available at: \emph{\small{\url{https://github.com/ghuawhu/end-to-end-synthetic-speech-detection}}}.

\begin{table}[!t]
	\renewcommand{\arraystretch}{1.0}
	\caption{EER (\%) of the proposed and state-of-the-art methods on ASVspoof2019 LA dev and eval sets, $M=4$, $C_\text{L}=\{64,32\}$, $C_\text{R}=\{32, 64, 128, 128\}$, $C_\text{I}=\{8,16,32,32\}$.}
	\label{main_results}
	\centering
	\vspace*{-6pt}
	\small{
	\begin{tabular}{r|c|c|c}
		\hline
		\hline
		\multicolumn{1}{c|}{{Method}}  & \#Param & \;Dev\; & \;Eval\;\\ 
		\hline
		\small{Baseline LFCC+GMM} \cite{ASVspoof2019_2} & - & $0.43$ & $9.57$ \\
		\small{Baseline CQCC+GMM}  \cite{ASVspoof2019_2} & - & $2.71$ & $8.09$\\
		\hline
		\small{Subband CQCC+MLP \cite{2020_Yang_TIFS_Subband_Det}}  & - & - & $8.04$\\
			\small{$8$ Features+MLP\cite{2019_Fuse_All}} & - & $0.00$& $4.13$\\
		\small{Spec+VGG+SincNet \cite{2019_VGG}} & \small{$>4.32$M} & $0.00$ & $8.01$\\
		\small{Spec+CQCC+ResNet+SE \cite{2019_SE_Res}} & $5.80$M & $0.00$& $6.70$\\
		\small{FFT+CNN \cite{2019_STC}} & $10.2$M & $0.04$ & $4.53$\\
		\small{$3$ Features+CNN \cite{2019_STC}} & $30.6$M & $0.00$ & $1.86$\\
		\small{CQT+Res2Net+SE \cite{2021_Li_ICASSP_Res2Net_Det}} & $0.92$M & $0.43$ & $2.50$ \\
		\small{$3$ Features+Res2Net+SE } \cite{2021_Li_ICASSP_Res2Net_Det} & $2.76$M  & $0.00$ & $1.89$\\
		\hline
		\small{CQT+2D-Res-TSSDNet} &$0.97$M & $0.59$ & $5.89$ \\
		\small{End-to-End Res-TSSDNet} & $0.35$M & $0.74$  & $\mathbf{1.64}$\\
		\small{End-to-End Inc-TSSDNet}  & $\mathbf{0.09}$M &  $1.09$ & $4.04$\\
		\hline
		\hline
	\end{tabular}
}
\end{table}

\subsection{Main Results}
The comparison of the results in terms of EER obtained on the logical access (LA) development and evaluation sets of the ASVspoof2019 challenge is presented in Table \ref{main_results}, where the 2D-Res-TSSDNet is the 2D version of the Res-TSSDNet, having the same architecture except that all the convolution and pooling ($2\times 2$ pooling) layers use 2D kernels instead.

\begin{table}[!t]
	\renewcommand{\arraystretch}{1.0}
	\caption{Ablation study of Res-TSSDNet and Inc-TSSDNet, using ASVspoof2019 LA eval EER (\%), $C_\text{L}=\{64, 32\}$.}
	\label{ablation}
	\centering
	\vspace*{-6pt}
	\setlength{\tabcolsep}{5pt}
	\small{
		\begin{tabular}{c|c|l|c|c|c}
			\hline
			\hline
			& $M$ & \multicolumn{1}{c|}{{ $C_\text{R}$}} & $1\times1$ & \#Param & Eval \\
			\hline
			\parbox[t]{3mm}{\multirow{5}{*}{\rotatebox[origin=c]{90}{\footnotesize{Res-TSSDNet}}}} & $3$ & $\{32, 64, 128\}$ & Yes & $0.18$M & $11.37$\\
			& $4$ & $\{32, 64,128,128\}$ & No & $0.32$M & $2.69$\\
			& $4$ & $\{32, 64,128,128\}$ &  Yes & $0.35$M & $\mathbf{1.64}$\\
			& $5$ & $\{32, 64,128,128, 128\}$ & No & $0.47$M & $5.14$\\
			& $5$ & $\{32, 64,128,128, 128\}$ & Yes & $0.51$M &$4.58$\\
		\end{tabular}
		\begin{tabular}{c|c|l|l|c|c}
			\hline
			\hline
			& $M$ & \multicolumn{1}{c|}{{ $C_\text{I}$}} & \multicolumn{1}{c|}{Dilation $d$}  & \#Param  & Eval \\
			\hline
			\parbox[t]{3mm}{\multirow{5}{*}{\rotatebox[origin=c]{90}{\footnotesize{Inc-TSSDNet}}}} & $3$ & $\{8, 16,32\}$ & $\{2^0,\ldots,2^3\}$ & $0.04$M & $10.39$\\
			& $4$ & $\{8, 16,32, 32\}$ &  $\{2^0,\ldots,2^3\}$ & $0.09$M & $4.04$\\
			& $5$ & $\{8, 16,32, 64, 64\}$ &  $\{2^0,\ldots,2^3\}$ & $0.35$M & $5.31$\\
			& $4$ & $\{8, 16,32, 32\}$ &  $\{2^0, \ldots, 2^7\}$ & $0.34$M & $\mathbf{3.75}$\\
			& $5$ & $\{8, 16,32, 64,64\}$ &  $\{2^0, \ldots, 2^7\}$ & $1.34$M & $4.20$\\
			\hline
			\hline
		\end{tabular}
	}
\end{table}

We make the following remarks from the main results. \textbf{i)} The works of \cite{2020_Yang_TIFS_Subband_Det} and \cite{2019_Fuse_All} represent the best results of sophisticated hand-crafted feature engineering plus an MLP as the back-end classifier. \textbf{ii)} The majority of recent works belong to the type of pre-transform (or light feature engineering) plus DNNs to further perform feature extraction and classification. \textbf{iii)} All the works incorporating DNNs rely on the feature and model fusion for performance improvement, and in \cite{2019_STC} and \cite{2021_Li_ICASSP_Res2Net_Det}, the fused results have achieved EERs below $2\%$. \textbf{iv)} The 2D-Res-TSSDNet result is obtained with experimental settings identical to \cite{2021_Li_ICASSP_Res2Net_Det} without fusion, and it can be seen that when working with 2D pre-transform input, the use of advanced DNN components, i.e., Res2Net and SE, becomes very necessary. \textbf{v)} Most importantly, the proposed Res-TSSDNet is a single end-to-end network (no fusion, no feature engineering), containing less than a half of trainable weights than the one in \cite{2021_Li_ICASSP_Res2Net_Det} and only about one-tenth than in \cite{2019_STC}, but it achieves the overall lowest evaluation EER by a clear margin. \textbf{vi)} Lastly, the Inc-TSSDNet is extremely light, having only $0.09$M parameters, but it could still achieve an EER lower than those from \cite{2019_SE_Res,2019_VGG,2019_STC} heavy models. 

\subsection{Ablation Study}
We first perform ablation study by varying the depth or width of the networks, and the results are summarized in Table \ref{ablation}. For the Res-TSSDNet, the column ``$1 \times 1$'' indicates whether the ``skip connection'' in Fig. \ref{models} (a) is used. It can be seen that going either shallower or deeper will result in the raise of EER, while with the use of ResNet skip connection, the network could achieve $1.05\%$ EER reduction over the one without using it. Similarly for the Inc-TSSDNet, the sweet spot also lies in the moderate depth or width.
 
\begin{figure}[!t]
	\centering
	\includegraphics[width=2.7in]{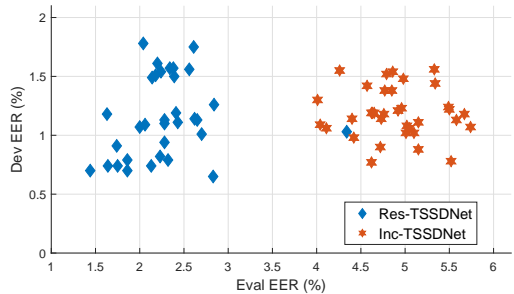}
	\vspace{-3pt}
	\caption{Intra-model performance on ASVspoof2019.}
	\label{sensitivity}
\end{figure}

We further perform intra-model sensitivity analysis using the two proposed end-to-end models in Table \ref{main_results}. Fixing all hyperparameters, the two models are trained from scratch using ASVspoof2019 training set for over $30$ times, and the dev and eval EERs are summarized in Fig. \ref{sensitivity}. It can be seen that the EERs of both models are bounded within certain ranges (except one outlier eval EER $>4\%$ for the Res-TSSDNet). The Inc-TSSDNet yields tighter dev EERs over the Res-TSSDNet, but eval EERs of the former are clearly higher. We can see from Fig. \ref{sensitivity} and Table \ref{ablation} that the intra-model differences may be as significant as the differences from model configurations. Relatively lighter models are hence recommended for their better trade-offs between accuracy and efficiency.

In addition, we have also discovered that \textbf{i)} changing all the activations from ReLU to leaky or parametric ReLU does not lead to a clear performance difference; \textbf{ii)} The first layer with a $1\times 7$ convolution kernel, adopted from ResNet setting, is slightly better than using a $1\times 3$ kernel; \textbf{iii)} Global max pooling is found to be more effective than global average pooling before the linear layers for both networks, but for the 2D-Res-TSSDNet, we stick to global average pooling; \textbf{iv)} The EERs of using standard CE are slightly higher than those using the WCE; \textbf{v)} Duration of training example also matters. Experimental results using $5$-second truncation yielded a slight performance degradation, but when $2$-second truncation is applied, the EER on evaluation set increased drastically.

\subsection{Cross-Dataset Testing}
\begin{figure}[!t]
	\centering
	\includegraphics[width=2.9in]{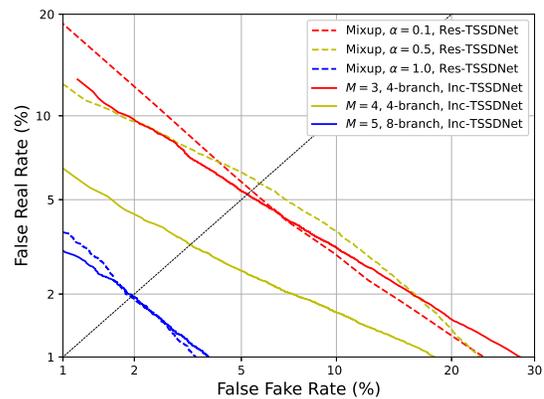}
	\vspace{-3pt}
	\caption{DET curves of cross-dataset testing on ASVspoof2015 eval set.}
	\label{ROC}
\end{figure}

We now perform the cross-dataset experiments. Since the ASVspoof2015 training set contains relatively old speech synthesis methods, we focus on using networks trained on the training set of more advanced ASVspoof2019 to test on the dev and eval sets of ASVspoof2015. The intra- and inter-dataset EERs are presented in Table \ref{cross_dataset}. It can be seen that the GMMs learned from LFCC and CQCC features in ASVspoof2019 are generally inconsistent with the data in ASVspoof2015. For the best Res-TSSDNet on ASVspoof2019, it could not generalize to ASVspoof2015 either, whose EERs indicate almost indistinguishable softmax probability distributions for real and fake classes. However, by incorporating mixup regularization and increasing the level of mixup level $\alpha$, we observe that the Res-TSSDNet can significantly reduce the cross dataset EERs to less than $2\%$, while slightly sacrificing the performance on the original dataset. Further, all the Inc-TSSDNets have very attractive generalization capability even for the lightest model. The $M=5$, $8$-branch version yields the best cross-dataset performance with $1.96\%$ eval EER. This is a significant score compared to the existing cross-dataset results as reported in \cite{Cross_Dataset_17,Cross_Dataset_19,Cross_Dataset_2020}. Noticeably in \cite{Cross_Dataset_2020}, also trained on ASVspoof2019 training set and tested on ASVspoof2015, the use of the CQT based features could only achieve EERs greater than $20\%$ (see Table 2 in \cite{Cross_Dataset_2020}). For completeness, the detection error trade-off (DET) curves on ASVspoof2015 evaluation set using a few methods in Table \ref{cross_dataset} are provided in Fig. \ref{ROC}.

\begin{table}[!t]
	\renewcommand{\arraystretch}{1.0}
	\caption{EER (\%) of networks trained on ASVspoof2019 training set, tested on ASVspoof2015 dev and eval sets.}
	\label{cross_dataset}
	\centering
	\vspace*{-6pt}
	\small{
		\begin{tabular}{r|c|c|c}
			\hline
			\hline
			\multicolumn{1}{c|}{\multirow{2}{*}{Method}}  & 2019 & \multicolumn{2}{c}{2015} \\
			\cline{2-4}
			& Eval  &  Dev & Eval \\
			\hline
			\small{Baseline LFCC+GMM} \cite{ASVspoof2019_2} & $9.57$  & $19.82$ & $15.91$ \\
			\small{Baseline CQCC+GMM}  \cite{ASVspoof2019_2} & $8.09$ &$47.72$ & $39.90$\\
			\hline
			\small{Res-TSSDNet}  & $1.64$ & $39.42$ & $42.52$\\
			\small{Mixup, $\alpha = 0.1$, Res-TSSDNet} & $2.07$ & $5.48$ & $5.46$\\
			\small{Mixup, $\alpha = 0.5$, Res-TSSDNet} & $2.29$ & $3.50$ & $5.75$\\
			\small{Mixup, $\alpha = 1.0$, Res-TSSDNet} & $2.16$ & $\mathbf{0.71}$ & $\mathbf{1.95}$\\
			\small{$M=3$, $4$-branch, Inc-TSSDNet} & $10.39$ & ${5.31}$ & ${5.24}$\\
			\small{$M=4$, $4$-branch, Inc-TSSDNet} & $4.04$ & ${2.78}$ & ${3.29}$\\
			\small{$M=4$, $8$-branch, Inc-TSSDNet} & $3.75$ & ${1.84}$ & $2.16$\\
			\small{$M=5$, $8$-branch, Inc-TSSDNet} & $4.20$ & $\mathbf{1.31}$ & $\mathbf{1.96}$\\
			\hline
			\hline
		\end{tabular}
	}
\end{table}

\section{Conclusion}
We have shown that a light-weight end-to-end neural network, significantly different from the exiting front- and back-end pipeline, could achieve to date the best synthetic speech detection results. It reduces the ASVspoof2019 eval EER by a clear margin compared to much heavier networks fed by pre-transform inputs, sophisticated hand-crafted features plus MLP classifiers, or the fusion of many systems of such kinds. We have further shown via cross-dataset testing that the proposed networks could also generalize to unseen dataset. In the ongoing ASVspoof2021 challenge, a new speech deepfake (DF) detection task is introduced specially for synthetic deepfake speech detection, and end-to-end methods are being given more attention, e.g., the RawNet2 \cite{2021_Tak_ICASSP_RawNet2} is used as a baseline.

\newpage
\bibliographystyle{IEEEtran}
\bibliography{ref_deepfake_audio}

\begin{thebibliography}{10}
\providecommand{\url}[1]{#1}
\csname url@samestyle\endcsname
\providecommand{\newblock}{\relax}
\providecommand{\bibinfo}[2]{#2}
\providecommand{\BIBentrySTDinterwordspacing}{\spaceskip=0pt\relax}
\providecommand{\BIBentryALTinterwordstretchfactor}{4}
\providecommand{\BIBentryALTinterwordspacing}{\spaceskip=\fontdimen2\font plus
\BIBentryALTinterwordstretchfactor\fontdimen3\font minus
  \fontdimen4\font\relax}
\providecommand{\BIBforeignlanguage}[2]{{%
\expandafter\ifx\csname l@#1\endcsname\relax
\typeout{** WARNING: IEEEtran.bst: No hyphenation pattern has been}%
\typeout{** loaded for the language `#1'. Using the pattern for}%
\typeout{** the default language instead.}%
\else
\language=\csname l@#1\endcsname
\fi
#2}}
\providecommand{\BIBdecl}{\relax}
\BIBdecl

\bibitem{2013_Tokuda_ProcIEEE_HMM}
K.~Tokuda, Y.~Nankaku, T.~Toda, H.~Zen, J.~Yamagishi, and K.~Oura, ``Speech
  synthesis based on hidden markov models,'' \emph{Proc. IEEE}, vol. 101,
  no.~5, pp. 1234--1252, May 2013.

\bibitem{2021_Ren_ICLR_FastSpeech2}
Y.~Ren, C.~Hu, X.~Tan, T.~Qin, S.~Zhao, Z.~Zhao, and T.-Y. Liu, ``{FastSpeech}
  2: Fast and high-quality end-to-end text to speech,'' in \emph{Proc. Int.
  Conf. Learning Representations (ICLR)}, 2021, pp. 1--15.

\bibitem{2017_Tian_TASLP_VC}
X.~{Tian}, S.~W. {Lee}, Z.~{Wu}, E.~S. {Chng}, and H.~{Li}, ``An exemplar-based
  approach to frequency warping for voice conversion,'' \emph{IEEE/ACM Trans.
  Audio, Speech, Lang. Process.}, vol.~25, no.~10, pp. 1863--1876, 2017.

\bibitem{2018_Gao_ICASSP_Impersonation}
Y.~Gao, R.~Singh, and B.~Raj, ``Voice impersonation using generative
  adversarial networks,'' in \emph{Proc. IEEE International Conference on
  Acoustics, Speech and Signal Processing (ICASSP 2018)}, Apr. 2018, pp.
  2506--2510.

\bibitem{2018_Arik_NIPS_Voice_Clone}
S.~{\"O}. Ar{\i}k, J.~Chen, K.~Peng, W.~Ping, and Y.~Zhou, ``Neural voice
  cloning with a few samples,'' in \emph{Proc. the 32nd International
  Conference on Neural Information Processing Systems (NeurIPS)}, 2018, pp.
  10\,019--10\,029.

\bibitem{2015_Sanchez_TIFS_Det}
J.~Sanchez, I.~Saratxaga, I.~Hern{\'a}ez, E.~Navas, D.~Erro, and T.~Raitio,
  ``Toward a universal synthetic speech spoofing detection using phase
  information,'' \emph{IEEE Trans. Inf. Forensics Security}, vol.~10, no.~4,
  pp. 810--820, Apr. 2015.

\bibitem{2015_Sahidullah_INTERSPEECH_Det}
M.~Sahidullah, T.~Kinnunen, and C.~Hanil\c{c}i, ``A comparison of features for
  synthetic speech detection,'' in \emph{Proc. Interspeech}, 2015.

\bibitem{2016_Sara_SC_Phase_Det}
I.~Saratxaga, J.~Sanchez, Z.~Wu, I.~Hernaez, and E.~Navas, ``Synthetic speech
  detection using phase information,'' \emph{Speech Communication}, vol.~81,
  pp. 30--41, 2016.

\bibitem{2017_Patel_JSTSP_Det}
T.~B. {Patel} and H.~A. {Patil}, ``Significance of source--filter interaction
  for classification of natural vs. spoofed speech,'' \emph{IEEE J. Sel. Topics
  Signal Process.}, vol.~11, no.~4, pp. 644--659, 2017.

\bibitem{2017_Patel_JSTSP_IF_Det}
------, ``Cochlear filter and instantaneous frequency based features for
  spoofed speech detection,'' \emph{IEEE J. Sel. Topics Signal Process.},
  vol.~11, no.~4, pp. 618--631, 2017.

\bibitem{2017_Paul_JSTSP_Det}
D.~Paul, M.~Pal, and G.~Saha, ``Spectral features for synthetic speech
  detection,'' \emph{IEEE J. Sel. Topics Signal Process.}, vol.~11, no.~4, pp.
  605--617, Jun. 2017.

\bibitem{2017_Wang_JSTSP_Phase_Det}
L.~{Wang}, S.~{Nakagawa}, Z.~{Zhang}, Y.~{Yoshida}, and Y.~{Kawakami},
  ``Spoofing speech detection using modified relative phase information,''
  \emph{IEEE J. Sel. Topics Signal Process.}, vol.~11, no.~4, pp. 660--670,
  2017.

\bibitem{2017_Todisco_CSL_CQT_Det}
M.~Todisco, H.~Delgado, and N.~Evans, ``Constant {Q} cepstral coefficients: A
  spoofing countermeasure for automatic speaker verification,'' \emph{Computer
  Speech \& Language}, vol.~45, pp. 516--535, 2017.

\bibitem{2018_Pal_CSL_hand_Det}
M.~Pal, D.~Paul, and G.~Saha, ``Synthetic speech detection using fundamental
  frequency variation and spectral features,'' \emph{Computer Speech \&
  Language}, vol.~48, pp. 31--50, 2018.

\bibitem{2019_Yang_TASLP_Octave_Det}
J.~{Yang}, R.~K. {Das}, and N.~{Zhou}, ``Extraction of octave spectra
  information for spoofing attack detection,'' \emph{IEEE/ACM Trans. Audio,
  Speech, Lang. Process.}, vol.~27, no.~12, pp. 2373--2384, Dec. 2019.

\bibitem{2016_Tian_APSIPA_CNN_Det}
X.~{Tian}, X.~{Xiao}, E.~S. {Chng}, and H.~{Li}, ``Spoofing speech detection
  using temporal convolutional neural network,'' in \emph{2016 Asia-Pacific
  Signal and Information Processing Association Annual Summit and Conference
  (APSIPA)}, 2016, pp. 1--6.

\bibitem{2017_Muck_TASLP_Hand_Det}
H.~{Muckenhirn}, P.~{Korshunov}, M.~{Magimai-Doss}, and S.~{Marcel},
  ``Long-term spectral statistics for voice presentation attack detection,''
  \emph{IEEE/ACM Trans. Audio, Speech, Lang. Process.}, vol.~25, no.~11, pp.
  2098--2111, 2017.

\bibitem{2018_Yu_TNNLS_Acoustic_Det}
H.~{Yu}, Z.~{Tan}, Z.~{Ma}, R.~{Martin}, and J.~{Guo}, ``Spoofing detection in
  automatic speaker verification systems using dnn classifiers and dynamic
  acoustic features,'' \emph{IEEE Trans. Neural Netw. Learn. Syst.}, vol.~29,
  no.~10, pp. 4633--4644, Oct. 2018.

\bibitem{2020_Yang_TIFS_Subband_Det}
J.~Yang, R.~K. Das, and H.~Li, ``Significance of subband features for synthetic
  speech detection,'' \emph{IEEE Trans. Inf. Forensics Security}, vol.~15, pp.
  2160--2170, 2020.

\bibitem{2017_Zhang_JSTSP_DL_Det}
C.~{Zhang}, C.~{Yu}, and J.~H.~L. {Hansen}, ``An investigation of deep-learning
  frameworks for speaker verification antispoofing,'' \emph{IEEE J. Sel. Topics
  Signal Process.}, vol.~11, no.~4, pp. 684--694, 2017.

\bibitem{2019_Fuse_All}
R.~K. Das, J.~Yang, and H.~Li, ``Long range acoustic features for spoofed
  speech detection,'' in \emph{Proc. Interspeech}, 2019, pp. 1058--1062.

\bibitem{2017_Chen_INTERSPEECH_ResNet_Det_Replay}
Z.~Chen, Z.~Xie, W.~Zhang, and X.~Xu, ``{ResNet} and model fusion for automatic
  spoofing detection,'' in \emph{Proc. Interspeech}, Aug. 2017, pp. 102--106.

\bibitem{2016_Qian_SC_Deep_Det}
Y.~Qian, N.~Chen, and K.~Yu, ``Deep features for automatic spoofing
  detection,'' \emph{Speech Communication}, vol.~85, pp. 43--52, 2016.

\bibitem{2020_Adiban_CSL_Siamese_Det}
M.~Adiban, H.~Sameti, and S.~Shehnepoor, ``Replay spoofing countermeasure using
  autoencoder and siamese networks on asvspoof 2019 challenge,'' \emph{Computer
  Speech \& Language}, vol.~64, no. 101105, pp. 1--13, 2020.

\bibitem{2017_Qian_TASLP_Deep_Det}
Y.~{Qian}, N.~{Chen}, H.~{Dinkel}, and Z.~{Wu}, ``Deep feature engineering for
  noise robust spoofing detection,'' \emph{IEEE/ACM Trans. Audio, Speech, Lang.
  Process.}, vol.~25, no.~10, pp. 1942--1955, 2017.

\bibitem{2019_Ensenble}
B.~Chettri, D.~Stoller, V.~Morfi, M.~A.~M. Ramírez, E.~Benetos, and B.~L.
  Sturm, ``{Ensemble Models for Spoofing Detection in Automatic Speaker
  Verification},'' in \emph{Proc. Interspeech}, 2019, pp. 1018--1022.

\bibitem{2019_SE_Res}
C.-I. Lai, N.~Chen, J.~Villalba, and N.~Dehak, ``{ASSERT: Anti-Spoofing with
  Squeeze-Excitation and Residual Networks},'' in \emph{Proc. Interspeech
  2019}, 2019, pp. 1013--1017.

\bibitem{2019_VGG}
H.~Zeinali, T.~Stafylakis, G.~Athanasopoulou, J.~Rohdin, I.~Gkinis, L.~Burget,
  and J.~Černocký, ``{Detecting Spoofing Attacks Using VGG and SincNet:
  BUT-Omilia Submission to ASVspoof 2019 Challenge},'' in \emph{Proc.
  Interspeech}, 2019, pp. 1073--1077.

\bibitem{2019_STC}
G.~Lavrentyeva, S.~Novoselov, A.~Tseren, M.~Volkova, A.~Gorlanov, and
  A.~Kozlov, ``{STC} antispoofing systems for the {ASVspoof2019} challenge,''
  in \emph{Proc. Interspeech}, 2019, pp. 1033--1037.

\bibitem{2021_Li_ICASSP_Res2Net_Det}
X.~Li, N.~Li, C.~Weng, X.~Liu, D.~Su, D.~Yu, and H.~Meng, ``Replay and
  synthetic speech detection with {Res2Net} architecture,'' in \emph{Proc. IEEE
  International Conference on Acoustics, Speech and Signal Processing (ICASSP
  2021)}, 2021.

\bibitem{ASVspoof2019_short}
X.~Wang and \emph{et al.}, ``{ASVspoof 2019}: A large-scale public database of
  synthesized, converted and replayed speech,'' \emph{Computer Speech \&
  Language}, vol.~64, no. 101114, pp. 1--24, 2020.

\bibitem{End2End_2019_Interspeech}
H.~Muckenhirn, V.~Abrol, M.~Magimai-Doss, and S.~Marcel, ``{Understanding and
  Visualizing Raw Waveform-Based CNNs},'' in \emph{Proc. Interspeech 2019},
  2019, pp. 2345--2349.

\bibitem{End2End_2019_ConvTasNet}
Y.~Luo and N.~Mesgarani, ``{Conv-TasNet}: Surpassing ideal time-frequency
  magnitude masking for speech separation,'' \emph{IEEE/ACM Trans. Audio,
  Speech, Lang. Process.}, vol.~27, no.~8, pp. 1256--1266, 2019.

\bibitem{2017_Muck_IJCB_E2E_Det}
H.~{Muckenhirn}, M.~{Magimai-Doss}, and S.~{Marcel}, ``End-to-end convolutional
  neural network-based voice presentation attack detection,'' in \emph{2017
  IEEE International Joint Conference on Biometrics (IJCB)}, 2017, pp.
  335--341.

\bibitem{2015_ResNet}
K.~He, X.~Zhang, S.~Ren, and J.~Sun, ``Deep residual learning for image
  recognition,'' \emph{arXiv, 1512.03385}, pp. 1--14, 2015.

\bibitem{2015_Inception}
C.~{Szegedy}, {Wei Liu}, {Yangqing Jia}, P.~{Sermanet}, S.~{Reed},
  D.~{Anguelov}, D.~{Erhan}, V.~{Vanhoucke}, and A.~{Rabinovich}, ``Going
  deeper with convolutions,'' in \emph{2015 IEEE Conference on Computer Vision
  and Pattern Recognition (CVPR)}, 2015, pp. 1--9.

\bibitem{ASVspoof2015}
Z.~Wu, T.~Kinnunen, N.~Evans, J.~Yamagishi, C.~Hanilci, M.~Sahidullah, and
  A.~Sizov, ``{ASVspoof 2015}: the first automatic speaker verification
  spoofing and countermeasures challenge,'' in \emph{Proc. Interspeech}, 2015,
  pp. 1--5.

\bibitem{2015_VGG}
K.~Simonyan and A.~Zisserman, ``Very deep convolutional networks for
  large-scale image recognition,'' in \emph{Proc. International Conference on
  Learning Representations (ICLR)}, 2015, pp. 1--14.

\bibitem{2016_Dilated_conv}
F.~Yu and V.~Koltun, ``{Multi-scale context aggregation by dilated
  convolutions},'' in \emph{Proc. International Conference on Learning
  Representations (ICLR)}, 2016, pp. 1--13.

\bibitem{2017_Adam}
D.~P. Kingma and J.~Ba, ``Adam: A method for stochastic optimization,'' in
  \emph{Proc. International Conference on Learning Representations (ICLR)},
  2017, pp. 1--15.

\bibitem{2018_mixup}
H.~Zhang, M.~Cisse, Y.~N. Dauphin, and D.~Lopez-Paz, ``Mixup: Beyond empirical
  risk minimization,'' in \emph{Proc. International Conference on Learning
  Representations (ICLR)}, 2018, pp. 1--13.

\bibitem{ASVspoof2019_2}
M.~Todisco, X.~Wang, V.~Vestman, M.~Sahidullah, H.~Delgado, A.~Nautsch,
  J.~Yamagishi, N.~Evans, T.~Kinnunen, and K.~A. Lee, ``{ASVspoof 2019}: Future
  horizons in spoofed and fake audio detection,'' in \emph{Proc. Interspeech},
  2019, pp. 1008--1012.

\bibitem{Cross_Dataset_17}
D.~Paul, M.~Sahidullah, and G.~Saha, ``Generalization of spoofing
  countermeasures: A case study with {ASVspoof} 2015 and {BTAS} 2016 corpora,''
  in \emph{IEEE International Conference on Acoustics, Speech and Signal
  Processing (ICASSP)}, 2017, pp. 2047--2051.

\bibitem{Cross_Dataset_19}
P.~Korshunov and S.~Marcel, ``A cross-database study of voice presentation
  attack detection,'' in \emph{Handbook of Biometric
  Anti-Spoofing\---Presentation Attack Detection, 2nd Ed.}\hskip 1em plus 0.5em
  minus 0.4em\relax Springer, 2019, pp. 363--389.

\bibitem{Cross_Dataset_2020}
R.~K. Das, J.~Yang, and H.~Li, ``Assessing the scope of generalized
  countermeasures for anti-spoofing,'' in \emph{IEEE International Conference
  on Acoustics, Speech and Signal Processing (ICASSP)}, 2020, pp. 6589--6593.

\bibitem{2021_Tak_ICASSP_RawNet2}
H.~Tak, J.~Patino, M.~Todisco, A.~Nautsch, N.~Evans, and A.~Larcher,
  ``End-to-end anti-spoofing with {RawNet2},'' in \emph{Proc. IEEE
  International Conference on Acoustics, Speech and Signal Processing
  (ICASSP)}, 2021, pp. 6369--6373.

\end{thebibliography}
\end{document}